\RequirePackage{fix-cm}

\documentclass[twocolumn,epjc3]{svjour3}  
\smartqed  
\RequirePackage{graphicx}
\usepackage{amssymb}
\usepackage{upgreek}
\usepackage{mathptmx}
\usepackage[english]{babel} 
\usepackage{xcolor}
\usepackage{ulem}
\usepackage{dcolumn}   
\usepackage{bm}        
\usepackage{amsbsy}   
\usepackage{multirow}
\usepackage{diagbox}
\usepackage{mathtools}
\usepackage{newtxtext}\usepackage[varvw]{newtxmath}
\usepackage{nicefrac}
\def \dif {\mathrm{d}}

\journalname{Eur. Phys. J. C}
\begin{document}

\title{Constraints on superheavy dark matter decaying into $h\nu$, $Z\nu$ and $W\ell$}
\subtitle{Benchmark example within an extended seesaw framework}


\author{Olivier Deligny\thanksref{e1,addr1}}

\thankstext{e1}{e-mail: deligny@ijclab.in2p3.fr}


\institute{Laboratoire de Physique des 2 Infinis Ir\`ene Joliot-Curie (IJCLab)
CNRS/IN2P3, Universit\'{e} Paris-Saclay, Orsay, France \label{addr1}}

\date{Received: date / Accepted: date}

\maketitle

\begin{abstract}
Dark matter particles could be superheavy (mass $M_X\gtrsim10^9~$GeV) provided that their lifetime $\tau_X$ is extremely long, i.e. greater than $\simeq 10^{22}~$yr. Such stringent constraints on $\tau_X$ are generally obtained by limiting the prompt emission of ultrahigh energy ($\gtrsim10^9~$GeV) gamma rays and neutrinos from the decay processes to below the corresponding flux upper bounds. In this paper, we show that even more severe bounds can be obtained for $M_X\gtrsim10^{13}~$GeV from the synchrotron radiation of electron decay byproducts in the Galaxy. We illustrate the power of these constraints using generic Higgs-induced  $h\nu$ and gauge-induced $Z\nu/W\ell$ decay channels, motivated by particle-physics setups invoking right-handed neutrinos. As a concrete benchmark, we consider a superheavy dark-matter candidate within an extended type-I seesaw framework and show that the lower bounds on lifetime can be translated into upper bounds on a mass-mixing parameter $\delta M$, which must satisfy approximately $\delta M\lesssim 2\times 10^{-17}/[M_X/(10^9~\mathrm{GeV})]^{0.5}$~GeV for $M_X\gtrsim 10^9$~GeV. Some implications in the context of inflationary cosmologies are discussed. 

\end{abstract}

\section{Introduction}
\label{sec:intro}

Dark matter (DM) remains elusive -- see e.g.~\cite{Cirelli:2024ssz} for a recent review. One possibility to explain its nature consists in considering that beyond-standard-model physics, including DM physics, only manifests itself at a very high energy scale. A particularly intriguing scale is around $10^{13}~$GeV. The mass of the inflaton, the scalar field responsible for the supposed inflationary phase in the early Universe, is indirectly inferred to range around $10^{13}~$GeV from the amplitude of the scalar perturbations observed in the cosmic microwave background~\cite{Planck:2018jri}. The mass of the right-handed neutrinos responsible for the mass generation of the active neutrinos is also in the range of $10^{13}~$GeV, at least within the vanilla seesaw mechanism~\cite{Minkowski:1977sc,Gell-Mann:1979vob,Yanagida1979,Mohapatra:1979ia}. Moreover, the instability scale of the Standard Model characterizing the scale at which the Higgs potential develops an instability at large field values is actually high, namely $10^{10}$-to-$10^{12}$\,GeV~\cite{Degrassi:2012ry,Alekhin:2012py,Bednyakov:2015sca}. Consequently, the Standard Model can be extrapolated to such high scales without the need to introduce beyond-standard-model physics to stabilize a potential that would suddenly become unbounded from below. The mass spectrum of the dark sector could also reflect this high-energy scale, as an intermediate and fundamental scale below Grand Unification and Planck ones. 

Superheavy particles of mass $M_X$ must be extremely long-lived  particles to be responsible for the relic density of DM observed today~\cite{Ellis:1990nb,Bhattacharjee:1999mup,Anchordoqui:2018qom}. The various generic constraints derived for the lifetime $\tau_X$ are particularly demanding: $\tau_X\gtrsim 10^{22}~$yr, see e.g.~\cite{Aloisio:2015lva,Ishiwata:2019aet,Guepin:2021ljb,Berat:2022iea,IceCube:2022clp,Das:2023wtk,PierreAuger:2023vql,Das:2024bed} for recent estimates in generic channels. These constraints are primarily derived from the non-observation of gamma-ray or neutrino fluxes at ultrahigh energies that would be suggestive of decay byproducts of superheavy particles. Independent of the specific decay channels at play, fragmentation processes indeed give rise to large fluxes of UHE particles such as nucleons, electrons, gamma rays and neutrinos. This is because of large radiative corrections at scales beyond the electroweak one as well as the triggering of QCD cascades until the hadronization of the partons occurs and the unstable hadrons eventually decay~\cite{Aloisio:2003xj,Sarkar:2001se,Barbot:2002gt,Kachelriess:2018rty,Alcantara:2019sco}. Among the decay byproducts, electrons trigger emission of secondary fluxes of gamma rays due to their radiation in the magnetic field of the Galaxy. The constraints that can be deduced from the latter emission process have long been neglected, and have only recently received some attention~\cite{Munbodh:2024ast}. After presenting in Section~\ref{sec:byproducts} the formalism used to infer constraints from both prompt and secondary emissions, we show in Section~\ref{sec:constraints} that the emission of secondary fluxes of gamma rays from the synchrotron radiation of electrons provides the most stringent constraints on the lifetime of DM particles for $M_X\simeq 10^{13}~$GeV. 

For a specific model to meet these constraints, it is generally necessary to fine-tune a reduced coupling constant $\alpha_X$ down to a tiny level and, simultaneously, the multiplicity of the final state to a large value~\cite{Hamaguchi:1998nj,deVega:2003hh,PierreAugerCollaboration:2022wir}. Interestingly, non-perturbative effects can effectively suppress the coupling constant and increase the multiplicity of final states~\cite{Kuzmin:1997jua,PierreAuger:2022jyk}. Alternatively, a handful theoretical constructions building upon the absence of right-handed degrees of freedom to describe neutrinos in the Standard Model can deliver superheavy DM candidates decaying into two- or three-body channels with viable $\tau_X$ based on motivated mechanisms lowering the coupling constant to the required level~\cite{Uehara:2001wd,Feldstein:2013kka,Dudas:2020sbq,Barman:2022qgt}. In Section~\ref{sec:model}, we explore a particular mass hierarchy within an extended type-I seesaw framework and show that for sufficiently small off-diagonal terms in their mass matrix, one of the superheavy-neutral-fermion mass eigenstate then fulfills the demanding ultra-long lifetime constraint. This benchmark model, which is shown to provide a viable DM candidate from particle-physics and cosmological aspects, illustrates the power of the lower bounds on lifetime derived in Section~\ref{sec:constraints} that directly translate into limits on the off-diagonal mass term.  General considerations about the benchmark model and the results are finally discussed in Section~\ref{sec:discussion}.

\section{Prompt and secondary decay byproducts}
\label{sec:byproducts}

If superheavy particles are disintegrating in the present-day Universe, their decay byproducts are necessarily, and quite independent of the specific decay channels, high and ultrahigh energy particles, including nucleons, electrons, neutrinos and gamma rays. This is due to fragmentation effects for particles with mass much larger than the electroweak scale. Seminal works in the QCD sector~\cite{Sarkar:2001se,Barbot:2002gt,Aloisio:2003xj} and in the electroweak one~\cite{Berezinsky:2002hq} have paved the way for the calculation of the fragmentation functions to derive the prompt flux of high-energy byproducts from the decay of a particle at high scale. Up-to-date fragmentation functions are available from the public code HDMSpectra~\cite{Bauer:2020jay}. The presence of electrons with energy reaching possibly $10^{13}~$GeV may also be relevant, as the magnetic field in the disk of the Galaxy is then strong enough to lead to ultrahigh energy gamma-rays by synchrotron radiation~\cite{Kachelriess:2018rty,Munbodh:2024ast}.

\subsection{Prompt emission}
\label{sec:prompt}

For any particle of ultrahigh energy $E$ of type $i$ (gamma-rays, (anti-)nucleons, neutrinos), the estimation of the diffuse flux (per steradian), $\phi_i(E,\mathbf{n})$, resulting from the prompt emission of decaying superheavy particles is obtained by integrating the position-dependent emission rate per unit volume and unit energy along the ``lookback position''  in the direction $\mathbf{n}$, 
\begin{equation}
    \label{eqn:flux_i}
    \phi_i(E,\mathbf{n})=\frac{1}{4\pi}\int_0^\infty \dif s~q_i(E,\mathbf{x}(s,\mathbf{n})).
\end{equation}
For neutral particles, the lookback position reduces to the position along the line of sight, $\mathbf{x}(s,\mathbf{n})=\mathbf{x}_\odot+s\mathbf{n}$, with $\mathbf{x}_\odot$ the position of the Solar system in the Galaxy and $\mathbf{n}\equiv\mathbf{n}(\ell,b)$ a unit vector on the sphere pointing to the longitude $\ell$ and latitude $b$, in Galactic coordinates. The attenuation of gamma-rays is negligible over galactic scales above $10^8~$GeV~\cite{Lipari:2018gzn}, which is here the threshold of interest. For (anti-)proton secondaries, the lookback position is obtained by retro-propagating particles from the Earth in the Galactic magnetic field. However, the deflections are negligible for energy thresholds above $\sim 10^{11}~$GeV, which will turn out to be those of interest in this study. Consequently, $\mathbf{x}(s,\mathbf{n})=\mathbf{x}_\odot+s\mathbf{n}$ is also used for (anti-)protons. 

The emission rate into the particle species $i$ is shaped by the DM density $n_\mathrm{DM}$, more conveniently expressed in terms of energy density $\rho_\mathrm{DM}=M_X\,n_\mathrm{DM}$, and by the differential decay partial widths $\dif\Gamma_{k,i}/\dif E=\Gamma_k\dif N_{k,i}/\dif E$ as
\begin{equation}
    \label{eqn:q_i}
    q_i(E,\mathbf{x})=\frac{\rho_\mathrm{DM}(\mathbf{x})}{M_X\tau_X}\sum_k \mathrm{BR}_k\frac{\dif N_{k,i}(E;M_X)}{\dif E},
\end{equation}
where $\mathrm{BR}_k$ stands for the branching ratio into the channel $k$. The ingredients are thus well separated in terms of particle-physics inputs, which regulate the spectra of secondaries from the decaying particle, and astrophysical factor. For the latter, there are uncertainties in the determination of the profile $\rho_\mathrm{DM}$. We use here the traditional NFW profile as a reference~\cite{Navarro:1995iw},
\begin{equation}
    \label{eqn:NFW}
    \rho_\mathrm{DM}(R)=\frac{\rho_s}{(R/R_s)(1+R/R_s)^2},
\end{equation}
where $R$ is the distance to the Galactic center, $R_s=24$\,kpc, and $\rho_s$ is fixed by the DM density in the solar neighborhood, namely $\rho_\odot\simeq0.44$\,GeV\,cm$^{-3}$~\cite{Nesti:2013uwa,Jiao:2023aci}. Other profiles such as those from Einasto~\cite{Einasto:1965czb}, Burkert~\cite{Burkert:1995yz} or Moore~\cite{Moore:1999nt} lead to differences on the order of a few percents in the constraints on the lifetime $\tau_X$  and subsequently on the mixing-angle parameters. Small contributions from extragalactic origin to the emission rate are neglected; the largest one concerns neutrinos and amount at most to 10\% of the Galactic one.

\subsection{Secondary emission of gamma-rays from synchrotron radiation}
\label{sec:secondary}

Electrons\footnote{Hereafter, ``electrons'' stand for electrons and positrons.} emitted from decay processes get rapidly attenuated in the Galaxy and do not reach the Earth. At the energies of interest, and by contrast with lower energy DM, secondary fluxes of ultrahigh energy gamma-rays from inverse Compton-scattering on low-energy photons from cosmic microwave, infra-red or star-light backgrounds are negligible compared to prompt fluxes. However, secondary fluxes from synchrotron emission can be significant for masses $M_X$ large enough. The emission rate entering into Eqn.~\ref{eqn:flux_i} is shaped this time by the (differential) synchrotron power $\dif P_\mathrm{syn}(E_\mathrm{e},E,\mathbf{x})/\dif E$ of electrons with energy $E_\mathrm{e}$ (and mass $m_\mathrm{e}$) emitting photons in the band between $E$ and $E+\dif E$ weighted by the (differential) density of electrons as~\cite{Cirelli:2009vg}
\begin{equation}
    \label{eqn:q_gamma_syn}
    q_\gamma(E,\mathbf{x})=\frac{1}{E}\int_{m_\mathrm{e}}^{M_X/2}\dif E_\mathrm{e}\frac{\dif n(E_\mathrm{e},\mathbf{x})}{\dif E_\mathrm{e}}\frac{\dif P_\mathrm{syn}(E_\mathrm{e},E,\mathbf{x})}{\dif E}.
\end{equation}
To a good approximation given that electrons with primary energy above $10^8~$GeV are attenuated over sub-parsec scales, the stationary density $\dif n/\dif E_\mathrm{e}$ results from a transport equation that retains only energy-loss and source terms,
\begin{equation}
    \label{eqn:transport_ne}
    \frac{\partial}{\partial E_\mathrm{e}} \left(b(E_\mathrm{e},\mathbf{x})\frac{\dif n(E_\mathrm{e},\mathbf{x})}{\dif E_\mathrm{e}} \right)=\frac{\rho_\mathrm{DM}(\mathbf{x})}{M_X\tau_X}\frac{\dif N_\mathrm{e}(E_\mathrm{e};M_X)}{\dif E_\mathrm{e}}.
\end{equation}
Here, $b(E_\mathrm{e},\mathbf{x})$ is the energy-loss rate due to synchrotron emission and inverse Compton scattering, which can be conveniently written as~\cite{Blumenthal:1970gc}
\begin{equation}
    \label{eqn:b}
    b(E_\mathrm{e},\mathbf{x})\simeq 1.03\times 10^{-16}\left(\frac{\rho_B(\mathbf{x})+\rho_\mathrm{r}}{\mathrm{eV~cm}^{-3}}\right)\left(\frac{E_\mathrm{e}}{\mathrm{GeV}}\right)^2~\mathrm{GeV~s}^{-1},
\end{equation}
with $\rho_\mathrm{r}\simeq 0.26~$eV~cm$^{-3}$ the radiation energy density due to the cosmic microwave background and $\rho_B$ that of the magnetic field $B(\mathbf{x})$,
\begin{equation}
    \label{eqn:rhoB}
    \rho_B(\mathbf{x})\simeq 2.5\times 10^{-2}\left(\frac{B(\mathbf{x})}{\upmu\mathrm{G}}\right)^2~\mathrm{eV~cm}^{-3}.
\end{equation}
The solution of Eqn.~\ref{eqn:transport_ne} reads as
\begin{equation}
    \label{eqn:ne}
     \frac{\dif n(E_\mathrm{e},\mathbf{x})}{\dif E_\mathrm{e}}=\frac{\rho_\mathrm{DM}(\mathbf{x})}{M_X\tau_Xb(E_\mathrm{e},\mathbf{x})}Y_\mathrm{e}(E_\mathrm{e}),
\end{equation}
with $Y_\mathrm{e}(E_\mathrm{e})=\int_{E_\mathrm{e}}^{M_X/2}\dif E'_\mathrm{e}~\dif N_\mathrm{e}(E'_\mathrm{e};M_X)/\dif E'_\mathrm{e}$ the yield of electrons with energy larger than $E_\mathrm{e}$. The synchrotron power, on the other hand, can conveniently be written as~\cite{Leite:2016lsv}
\begin{equation}
    \label{eqn:p_syn}
    \frac{\dif P_\mathrm{syn}(E_\mathrm{e},E,\mathbf{x})}{\dif E}=\frac{b_B(E_\mathrm{e},\mathbf{x})}{E_\mathrm{c}(E_\mathrm{e},\mathbf{x})}\Tilde{F}\left(\frac{E}{E_\mathrm{c}(E_\mathrm{e},\mathbf{x})}\right),
\end{equation}
with $b_B(E_\mathrm{e},\mathbf{x})$ the energy-loss rate due to synchrotron emission only, $E_\mathrm{c}$ a critical energy defined as
\begin{equation}
    \label{eqn:Ec}
    E_\mathrm{c}(E_\mathrm{e},\mathbf{x})\simeq 6.61\times 10^{-17}\left(\frac{E_\mathrm{e}}{\mathrm{GeV}}\right)^2\left(\frac{B_\perp(\mathbf{x})}{\upmu\mathrm{G}}\right)~\mathrm{GeV},
\end{equation}
and $\tilde{F}(x)$ defined as
\begin{equation}
    \label{eqn:F}
    \tilde{F}(x)=\frac{9\sqrt{3}}{8\pi}x\int_x^\infty\dif x'K_{5/3}(x')
\end{equation}
in the case of a uniform magnetic field~\cite{Longair:1992ze} (where $K_z(x)$ is the modified Bessel function of order $z$), and 
\begin{equation}
    \label{eqn:F}
    \tilde{F}(x)=\frac{27\sqrt{3}}{4\pi}x^2\left(K_{4/3}(x)K_{1/3}(x)-\frac{3x}{5}(K^2_{4/3}(x)-K^2_{1/3}(x))\right)
\end{equation}
in the case of a turbulent magnetic field~\cite{Ghisellini:1996fq}. Despite recent progresses, the knowledge of the Galactic magnetic field remains uncertain. We follow \cite{Jansson:2012pc} for the coherent component and consider a Kolmogorov turbulence with an intensity scaled to the coherent component. To bracket the systematic uncertainties in the lifetime $\tau_X$ inherited from those of the magnetic field, a factor three will be considered for the strength of the turbulence compared to the reference one.

\section{Generic constraints on lifetime}
\label{sec:constraints}

We derive in this section lower bounds on the lifetime\footnote{The DM candidate is denoted as $X$ in this section.} $\tau_X$ of superheavy DM from the existing constraints on fluxes of ultrahigh energy particles above $10^8~$GeV. Rather than setting these limits for a wide range of generic channels such as $e^+e^-$, $\nu\bar{\nu}$, $b\bar{b}$, $W^+W^-$, etc., we restrict ourselves to electroweak two-body channels $h\nu$, $Z\nu$ and $We$, which are of interest for specific models based, for instance, on metastable gravitinos in the context of supersymmetry broken at high scale with tiny $R$-parity violations~\cite{Feldstein:2013kka,Dudas:2019gkj,Addazi:2017kbx}, or on a Majorana mass term for the right handed part of a singlet Dirac fermion that propagates in the bulk of a 5D spacetime~\cite{Uehara:2001wd,Feldstein:2013kka}, or on one of the three right-handed neutrinos in the type-I seesaw framework~\cite{Barman:2022qgt}. This approach using specific channels is preferred here to that, widely-used in the literature, using the generic channels forementioned, which hardly meet concrete models. Another setup based on an extended seesaw framework that yields to two-body electroweak channels  will be presented in Section~\ref{sec:model}.

The expected number of ultrahigh energy particle $i$ produced as a direct decay byproduct or by fragmentation for any given observatory with directional exposure $\omega_i(E,\mathbf{n})$ reads as
\begin{equation}
    \label{eqn:n}
    \mu_i(>E) = \int_{4\pi}\dif\mathbf{n}\int_{>E}\dif E'~\omega_i(E',\mathbf{n})\phi_i(E',\mathbf{n}).
\end{equation}
We use the following most stringent constraints from the  contemporary Pierre Auger~\cite{PierreAuger:2015eyc}, Telescope Array~\cite{TelescopeArray:2012uws} and IceCube~\cite{IceCube:2016zyt} observatories:
\begin{itemize}
    \item No gamma-ray fluxes have been uncovered above $10^{7.3}~$GeV until now. The best sensitivity above that threshold is provided by the Pierre Auger Observatory~\cite{PierreAuger:2023nkh,PierreAuger:2022uwd,PierreAuger:2024ayl,PierreAuger:2022aty}. Because there is an irreducible small number of charged cosmic-ray events that resemble gamma-ray events despite the selection process, gamma-ray searches are not background-free. The values used for $\mu_\gamma(>E)$, given in Ref.~\cite{PierreAuger:2023vql}, result from the upper limits on the number of candidates accounting for this irreducible background. 
    \item No neutrinos have been observed above $10^8~$GeV at the IceCube~\cite{IceCube:2018fhm} and Pierre Auger~\cite{PierreAuger:2019ens} observatories, so that lower limits on $\tau_X$ can be inferred from Eqn.~\ref{eqn:n} by setting $\mu_\nu(>E)=2.39$~\cite{Feldman:1997qc}.
    \item No cosmic rays, including protons, have been observed above $10^{11.2}~$GeV at both the Pierre Auger and Telescope Array observatories, so that lower limits on $\tau_X$ can be inferred from Eqn.~\ref{eqn:n} by setting $\mu_\nu(>E)=2.39$~\cite{Feldman:1997qc}.
\end{itemize}
The up-to-date directional exposure of the Pierre Auger Observatory to gamma-rays and neutrinos is taken from~\cite{PierreAuger:2023vql}, while that to cosmic rays is taken from~\cite{PierreAuger:2022axr}. On the other hand, the directional exposure of the IceCube Observatory to through-going muon-neutrinos from the Northern Hemisphere is inferred from the directional-dependent effective area reported in~\cite{Stettner:2019tok}, which provides an absolute calibration at $10^8~$GeV enabling calculation of the directional-dependent Earth opacity to neutrinos with energies above $10^8~$GeV.

\begin{figure}[t]
\centering
\includegraphics[width=0.5\textwidth]{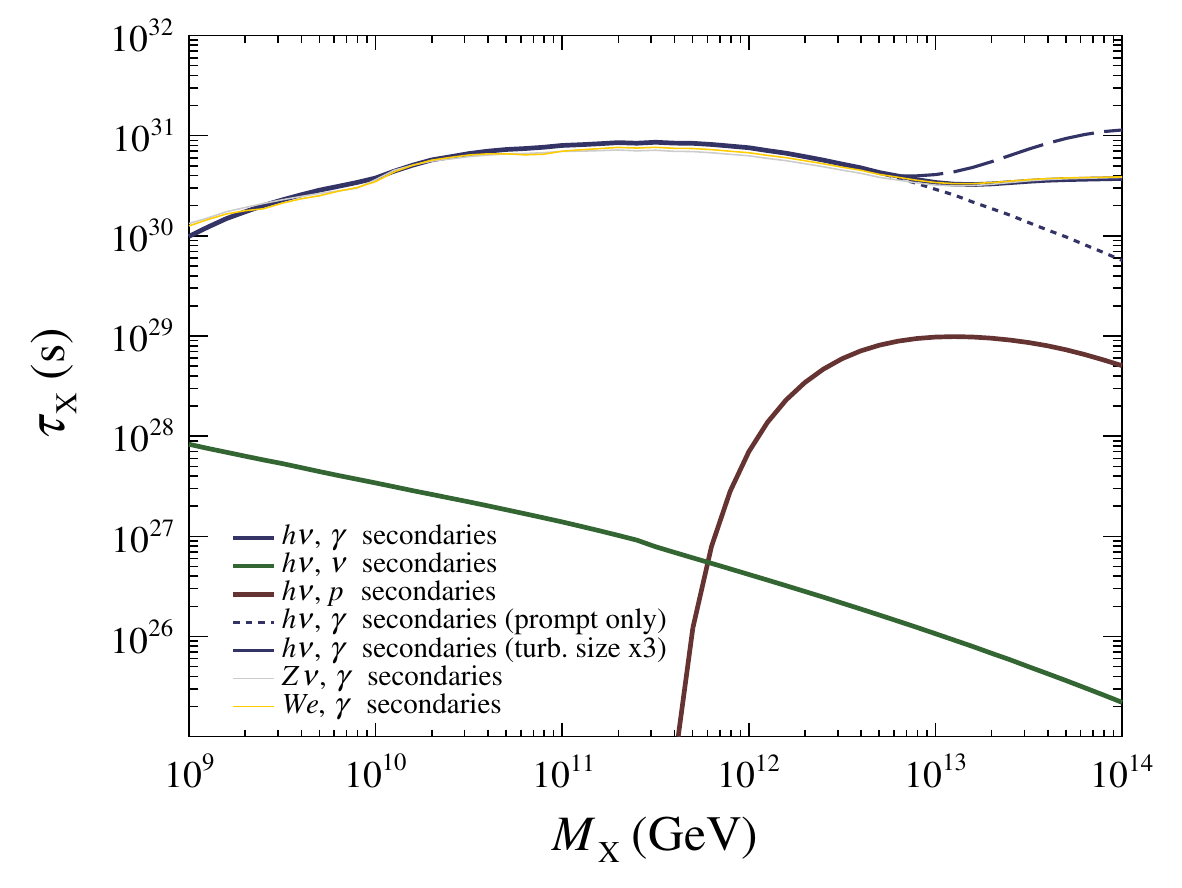}
\caption{Constraints on the lifetime $\tau_X$ from electroweak two-body decay channels as a function of $M_X$.}
\label{fig:tauX}
\end{figure}

The differential decay spectrum into particles of type $i$, $\dif N_i/\dif E$, results from the convolution of the differential decay width of $X$ to an intermediate particle $j$, $\dif \Gamma_j^X/\dif E$, with the fragmentation function $D_i^j(x,Q)$ of $j$ into $i$~\cite{Sarkar:2001se},
\begin{equation}
    \label{eqn:spectrum}
    \frac{\dif N_i(E;M_X)}{\dif E}=\frac{2}{M_X\Gamma_X}\sum_j\int_x^1\frac{\dif y}{y}\frac{\dif \Gamma_j(y)}{\dif y}D_i^j\left(\tfrac{x}{y},\tfrac{M_X}{2}\right),
\end{equation}
with $x=2E/M_X$. The first argument of the fragmentation functions is the fraction of energy of the created particle in units of the high scale described by the second argument. For a two-body channel $X\rightarrow ab$, this reduces to
\begin{equation}
    \label{eqn:spectrum-2body}
    \frac{\dif N_i(E;M_X)}{\dif E}=\frac{2}{M_X}\sum_{j=a,b}D_i^j\left(\tfrac{2E}{M_X},\tfrac{M_X}{2}\right).
\end{equation} 

Lower limits on the lifetime $\tau_X$ are obtained by setting, for a specific value of $M_X$, $n_i(E)$ to the 90\% confidence level upper-limit numbers corresponding to the number of background-event candidates in the absence of signal. Subsequently, a scan in $M_X$ is carried out. It leads to a curve in the plane $(\tau_{X},M_X)$ that pertains to the energy threshold $E$ considered. By repeating the procedure for several thresholds, a set of curves is obtained, reflecting the sensitivity of a specific energy threshold to some range of mass $M_X$. The union of the excluded regions finally provides the constraints in the $(\tau_{X},M_X)$ plane. Results are shown in Fig.~\ref{fig:tauX}; values below the lines are excluded. To illustrate the contribution from each secondary at our disposal, we show as the blue (red) [green] line the constraints obtained from gamma-rays (protons) [neutrinos] in the $h\nu$ channel. Constraints from gamma-rays turn out to be the most stringent in the mass range considered. The dotted line, visible above a few $10^{12}~$GeV where the synchrotron emission from electrons becomes relevant, is obtained from the prompt emission alone while, to illustrate the systematic uncertainties stemming from the magnetic field in the Galaxy, the long-dotted line is obtained by amplifying the contribution of the turbulent component by a factor three. Finally, the constraints from gamma-rays pertaining to the $Z\nu$ and $We$ channels are shown as the gray and orange lines, respectively; as a result of the fragmentation processes, they are almost indistinguishable from that of the $h\nu$ channel. Note that a generic branching ratio $\mathrm{BR}_k=1/3$ has been assumed throughout this section.

\section{Benchmark of superheavy dark matter within an extended seesaw framework}
\label{sec:model}

As a concrete example of superheavy DM model, we consider in this section an extended type-I seesaw framework, already explored in~\cite{Rott:2014kfa}. In this framework, superheavy right-handed neutrinos  provide the mass to the known neutrinos through the seesaw mechanism while another superheavy neutral lepton plays a key role for producing a Majorana DM particle. The phenomenology built from renormalizable terms alone that yields to the electroweak two-body decay of the DM candidate through a tiny mixing in the mass sector, already explored in~\cite{Rott:2014kfa}, is first presented in Section~\ref{sec:seesawmodel}. Second, the constraints on the lifetime are translated in terms of constraints on the mixing angle in Section~\ref{sec:mixing}. Implications of the tiny viable values for the mixing angle in terms of a UV completion model are also presented. Finally, the viable range of cosmological parameters for producing the right amount of the DM candidate in the early Universe is given in Section~\ref{sec:cosmology}.

\subsection{Metastable dark matter candidate within an extended seesaw framework}
\label{sec:seesawmodel}

We consider an extended type-I seesaw framework in which a superheavy neutral lepton, $\chi_\mathrm{R}$, has interactions with Standard Model particles suppressed by a mixing angle $\theta_\chi^2$.\footnote{In this benchmark, the realization of the DM scenario is possible for each flavor family. Consequently, only one flavor family is considered in this section; this simplification enables us to concentrate on the DM aspects of this benchmark.} $N_\mathrm{R}$ provides the mass to the known neutrino through the seesaw mechanism. Under these conditions, there is no explicit interactions of $\chi_\mathrm{R}$ in the basis of flavor eigenstates and the Majorana sector of the Lagrangian in this basis reads as follows,
\begin{equation}
    \label{eqn:LM}
    \mathcal{L}_\mathrm{M}=-\left(\frac{1}{2}M_\chi\overline{{\chi^\mathrm{c}}_{\mathrm{L}}}\chi_\mathrm{R}+\frac{1}{2}M_N\overline{{N^\mathrm{c}}_{\mathrm{L}}}N_\mathrm{R}+\delta M~ \overline{{\chi^\mathrm{c}}_{\mathrm{L}}}N_\mathrm{R}+~\mathrm{h.c.}\right),
\end{equation}
where $\chi^\mathrm{c}$ and $N^\mathrm
c$ are the field conjugates of the neutral fermions, $M_\chi$ and $M_N$ are the Majorana masses of the neutral fermions, $\delta M$ is a tiny mixing-mass parameter, and h.c. stands for hermitian conjugate. On the other hand, the fulcrum of the Dirac mass term is from the Standard-Model gauge invariant and renormalizable operator $\mathcal{O}=y_N\overline{L}\tilde{H}N_\mathrm{R}+y_\chi\overline{L}\tilde{H}\chi_\mathrm{R}+~\mathrm{h.c.}$, where $\tilde{H}\equiv \mathrm{i}\sigma_2H^\star$ is the dual of the Higgs isodoublet of the Standard Model $H=(h^+~h^0)^\mathrm{T}$, with $\sigma_2$ the second Pauli matrix. After electroweak spontaneous symmetry breaking, $h^0$ (and $h^{0\star}$) acquires the vacuum expectation value $v/\sqrt{2}$ and the operator $\mathcal{O}$ provides the Dirac mass terms
\begin{equation}
    \label{eqn:LD}
    \mathcal{L}_\mathrm{D}=-\left(\frac{y_Nv}{\sqrt{2}}\overline{\nu_\mathrm{L}}N_\mathrm{R}+\frac{y_\chi v}{\sqrt{2}}\overline{\nu_\mathrm{L}}\chi_\mathrm{R}+~\mathrm{h.c.}\right).
\end{equation} 
On introducing the vector $\mathcal{N}_\mathrm{R}=({\nu^\mathrm{c}}_\mathrm{R}~\chi_\mathrm{R}~N_\mathrm{R})^\mathrm{T}$, both $\mathcal{L}_\mathrm{M}$ and $\mathcal{L}_\mathrm{D}$ can be conveniently written in terms of a single mass term,
\begin{equation}
    \label{eqn:Lmass}
    \mathcal{L}_\mathrm{mass}=-\frac{1}{2}
	\overline{{\mathcal{N}^\mathrm{c}}_\mathrm{L}} \mathcal{M} \mathcal{N}_\mathrm{R}
    +~\mathrm{h.c.},
\end{equation}
where the matrix $\mathcal{M}$ is defined as
\begin{equation}
    \label{eqn:M}
    \mathcal{M}=
	\begin{pmatrix}
		0 & \delta m_\mathrm{D} & m_\mathrm{D} \\
		\delta m_\mathrm{D} & M_\chi & \delta M \\
        m_\mathrm{D} & \delta M & M_N
	\end{pmatrix},
\end{equation}
with $m_\mathrm{D}=y_Nv/\sqrt{2}$ and $\delta m_\mathrm{D}=y_\chi v/\sqrt{2}$. We assume the mass hierarchy to be such that 
\begin{equation}
    \delta m_\mathrm{D}\ll \delta M\ll  m_\mathrm{D}\ll M_\chi<M_N,
\end{equation}
which will be justified in Section~\ref{sec:mixing}. Adopting $\delta m_\mathrm{D}=0$ in the following, $\mathcal{M}$ can be written in terms of the product $\mathcal{M}=\mathcal{P}\mathcal{D}\mathcal{P}^{-1}$, 
\begin{equation}
    \label{eqn:M}
    \mathcal{M}=
    \begin{pmatrix}
	-\mathrm{i} & 0 & \theta_N \\
    0 & 1 & \theta_\chi \\
    -\mathrm{i}\theta_N & -\theta_\chi & 1
	\end{pmatrix}
    \begin{pmatrix}
	|m_{\tilde{\nu}}| & 0 & 0 \\
    0 & M_{\tilde{\chi}} & 0 \\
    0 & 0 & M_{\tilde{N}}
	\end{pmatrix}
    \begin{pmatrix}
	-\mathrm{i} & 0 & -\mathrm{i}\theta_N \\
    0 & 1 & -\theta_\chi \\
    \theta_N & \theta_\chi & 1
	\end{pmatrix},
\end{equation}
where the mixing angles $\theta_N=m_\mathrm{D}/M_N$ and $\theta_\chi=\delta M/\Delta M$ have been introduced, with $\Delta M=M_N-M_\chi$. The three eigenvalues of $\mathcal{M}$ are $m_{\tilde{\nu}}\simeq -m_\mathrm{D}^2/M_N$, $M_{\tilde{\chi}}\simeq M_\chi$ and $M_{\tilde{N}}\simeq M_N$. Note that we have absorbed the negative sign of $m_{\tilde{\nu}}$ through a redefinition of the some phases in the change-of-basis matrix $\mathcal{P}$. In this manner, the eigenvectors $\tilde{\mathcal{N}}_\mathrm{R}$ and $\tilde{\mathcal{N}}^\mathrm{c}_\mathrm{~~L}$ are 
\begin{equation}
    \label{eqn:tildeR}
    \tilde{\mathcal{N}}_\mathrm{R}=
    \begin{pmatrix}
	\tilde{\nu}_\mathrm{R} \\
    \tilde{\chi}_\mathrm{R} \\
    \tilde{N}_\mathrm{R}
	\end{pmatrix} =
    \begin{pmatrix}
	-\mathrm{i}{\nu^\mathrm{c}}_{\mathrm{R}} - \mathrm{i}\theta_NN_{\mathrm{R}} \\
    \chi_{\mathrm{R}}-\theta_\chi N_{\mathrm{R}}\\
    \theta_N{\nu^\mathrm{c}}_{\mathrm{R}}+\theta_\chi \chi_{\mathrm{R}}+N_{\mathrm{R}}
	\end{pmatrix}
\end{equation}
and 
\begin{equation}
    \tilde{\mathcal{N}}^\mathrm{c}_\mathrm{~~L}=
    \label{eqn:tildeL}
    \begin{pmatrix}
	{\tilde{\nu}}^\mathrm{c}_{\mathrm{~L}} \\
    {\tilde{\chi}^\mathrm{c}}_{\mathrm{~L}} \\
    {\tilde{N}^\mathrm{c}}_{\mathrm{~L}}
	\end{pmatrix} =
    \begin{pmatrix}
	-\mathrm{i}\nu_{\mathrm{L}} - \mathrm{i}\theta_N{N^\mathrm{c}}_{\mathrm{L}} \\
    {\chi^\mathrm{c}}_{\mathrm{L}}-\theta_\chi{N^\mathrm{c}}_{\mathrm{L}}\\
    \theta_N\nu_{\mathrm{L}}+\theta_\chi {\chi^\mathrm{c}}_{\mathrm{L}}+{N^\mathrm{c}}_{\mathrm{L}}
	\end{pmatrix},
\end{equation}
where terms proportional to $\theta_\chi^2$, $\theta_N^2$ and  $\theta_N\theta_\chi$ have been neglected. By defining the Majorana states $\tilde{\mathcal{N}}$ such that
\begin{equation}
    \label{eqn:tildestates}
    \tilde{\mathcal{N}}=\tilde{\mathcal{N}}^\mathrm{c}_\mathrm{~~L}+\tilde{\mathcal{N}}_\mathrm{R}=
    \begin{pmatrix}
	\tilde{\nu} \\
    \tilde{\chi} \\
    \tilde{N} 
	\end{pmatrix} =
    \begin{pmatrix}
	{\tilde{\nu}^\mathrm{c}}_{\mathrm{~L}}+\tilde{\nu}_{\mathrm{R}} \\
    {\tilde{\chi}^\mathrm{c}}_{\mathrm{~L}}+\tilde{\chi}_{\mathrm{R}} \\
    {\tilde{N}^\mathrm{c}}_{\mathrm{~L}}+\tilde{N}_{\mathrm{R}} 
	\end{pmatrix},
\end{equation}
it is then convenient to summarize the mass sector of the neutral-fermion Lagrangian as 
\begin{equation}
    \label{eqn:Lmass-diag}
    \mathcal{L}_\mathrm{mass}=-\frac{1}{2}
	\overline{\tilde{\mathcal{N}}} \mathcal{D}	\mathcal{\tilde{N}}.
\end{equation}

\subsection{Viable ranges of mixing angles and implications}
\label{sec:mixing}

\begin{table*}
\caption{Charge assignment, $Q$, of the UV particle content for each of the high-scale global $U(1)$ symmetry.}
\label{tab:Q}
\begin{tabular}{l c c c c c c c c c c c c c}
\hline\noalign{\smallskip}
 & $L$ & $\chi_\mathrm{R}$ & $N_\mathrm{R}$ & $\phi_\chi$ & $\phi_N$ & $\psi_1$ & $\tilde{\psi}_1$ & $\psi_2$ & $\tilde{\psi}_2$ & $\psi_3$ & $\tilde{\psi}_3$ & $\psi_4$ & $\tilde{\psi}_4$ \\
\noalign{\smallskip}\hline\noalign{\smallskip}
$Q$ & $-1$ & $-7$ & $+1$ & $+14$ & $-2$ & $+1$ & $-1$ & $-13$ & $+13$ & $-11$ & $+11$ & $-9$ & $+9$ \\
\noalign{\smallskip}\hline
\end{tabular}
\end{table*}

\begin{figure}[t]
\centering
\includegraphics[width=0.5\textwidth]{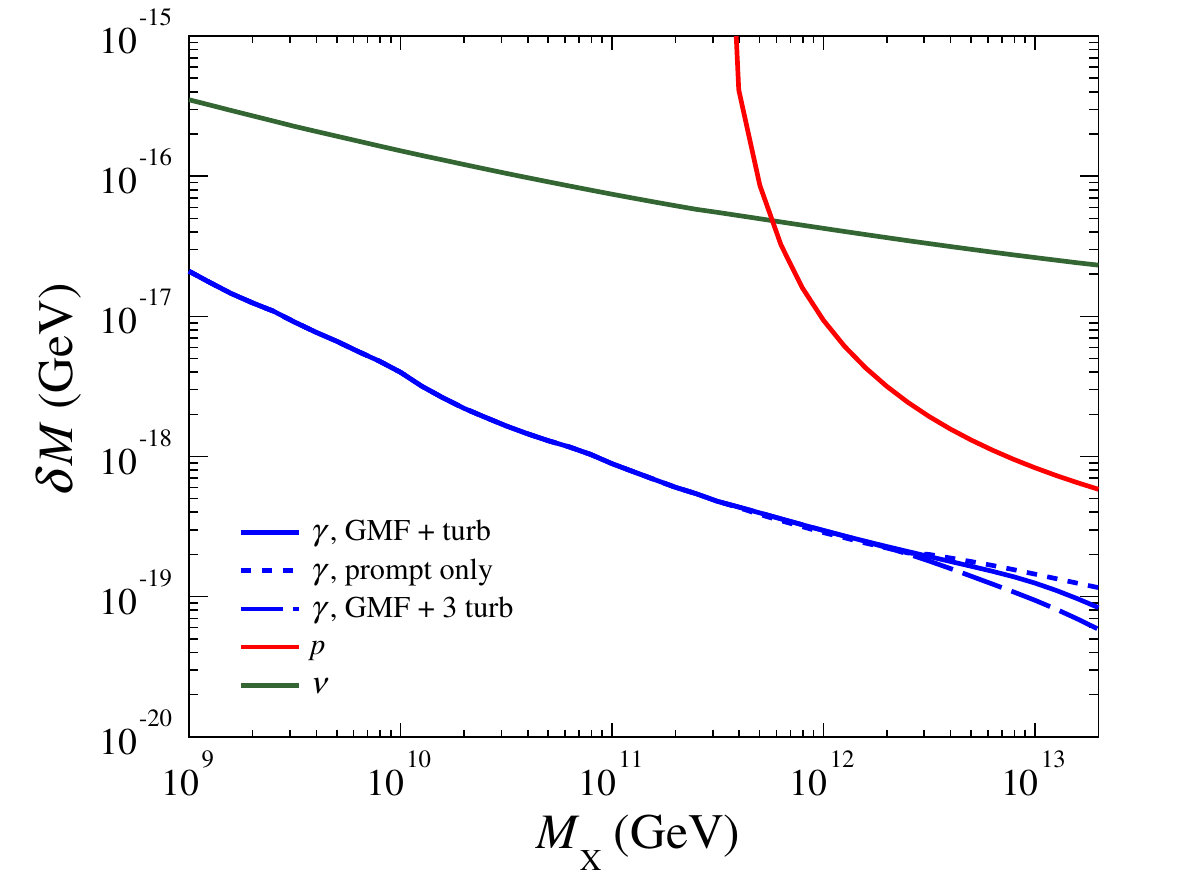}
\caption{Constraints on the mass-mixing parameter $\delta M$ as a function of $M_\chi$.}
\label{fig:dM}
\end{figure}
Despite the absence of explicit interactions of $\chi_\mathrm{R}$ with Standard Model particles, the small admixture of ${\chi^\mathrm{c}}_\mathrm{L}+\chi_\mathrm{R}$ contained in the state $\tilde{N}$, controlled by $\theta_\chi$, provides a decay channel for $\tilde{\chi}$. To see that, let's express the flavor eigenstates as a function of the mass eigenstates to first order in $\theta_\chi$ and $\theta_N$, 
\begin{equation}
    \label{eqn:tildeR-inv}
    \begin{pmatrix}
	{\nu^\mathrm{c}}_\mathrm{R} \\
    \chi_\mathrm{R} \\
    N_\mathrm{R}
	\end{pmatrix} =
    \begin{pmatrix}
	\mathrm{i}\tilde{\nu}_{\mathrm{R}} + \theta_N\tilde{N}_{\mathrm{R}} \\
    \tilde{\chi}_{\mathrm{R}} + \theta_\chi\tilde{N}_{\mathrm{R}}\\
    -\mathrm{i}\theta_N\tilde{\nu}_{\mathrm{R}}-\theta_\chi \tilde{\chi}_{\mathrm{R}}+\tilde{N}_{\mathrm{R}}
	\end{pmatrix},
\end{equation}
\begin{equation}
    \label{eqn:tildeL-inv}
    \begin{pmatrix}
	\nu_\mathrm{L} \\
    {\chi^\mathrm{c}}_\mathrm{L} \\
    {N^\mathrm{c}}_\mathrm{L}
	\end{pmatrix} =
    \begin{pmatrix}
	\mathrm{i}{\tilde{\nu}^\mathrm{c}}_{\mathrm{~L}} + \theta_N{\tilde{N}^\mathrm{c}}_{\mathrm{~L}} \\
    {\tilde{\chi}^\mathrm{c}}_{\mathrm{~L}} +\theta_\chi{\tilde{N}^\mathrm{c}}_{\mathrm{~L}}\\
    -\mathrm{i}\theta_N{\tilde{\nu}^\mathrm{c}}_{\mathrm{~L}} -\theta_\chi {\tilde{\chi}^\mathrm{c}}_{\mathrm{~L}}+{\tilde{N}^\mathrm{c}}_{\mathrm{~L}}
	\end{pmatrix},
\end{equation}
where the Majorana properties of $\tilde{N}$ are used (e.g. $\tilde{N}^\mathrm{c}=\tilde{N}$, so that ${\tilde{N}_\mathrm{R}}^\mathrm{~c}={\tilde{N}^\mathrm{c}}_\mathrm{~L}=\tilde{N}_\mathrm{L}$). After spontaneous symmetry breaking, the neutral scalar $h^0$ fluctuates about its vacuum expectation value, $h^0=(v+h)/\sqrt{2}$, with $h$ the Higgs boson. The operator $\mathcal{O}$ contains then a term that allows $N_\mathrm{R}$ to decay into $h\nu$. Once expressed in the mass-eigenstate basis, the decay channel $\tilde{\chi}\rightarrow h\tilde{\nu}$ appears,\footnote{Note that in this particular setup, the partial widths for the gauge-induced channels $\tilde{\chi}\rightarrow Z\tilde{\nu}$ and $\tilde{\chi}\rightarrow We$ are additionally suppressed by $\theta_N^2$.} controlled by $y_N\theta_\chi$: 
\begin{equation}
    \label{eqn:Ldecay}
    \frac{y_N}{\sqrt{2}}h\left(\overline{\nu_\mathrm{L}}N_\mathrm{R}+\overline{N_\mathrm{R}}\nu_\mathrm{L}\right)=\dots+ \frac{\mathrm{i}y_N\theta_\chi}{2\sqrt{2}}h\left(\overline{\tilde{\nu}}\tilde{\chi}+\overline{\tilde{\chi}}\tilde{\nu}\right)+\dots
\end{equation}
This is the result sought for. This interaction term gives rise to the width
\begin{equation}
    \label{eqn:decaywidth}
    \Gamma_{\tilde{\chi}\rightarrow h{\tilde{\nu}}}=\frac{\left|y_N\right|^2\theta_\chi^2}{256\pi}M_{\tilde{\chi}}
\end{equation}
for the DM candidate $\tilde{\chi}$. For $M_\chi$ not too close to $M_N$, a tiny mixing mass term $\delta M$ is thus enough to an extremely short (long) width (lifetime). Assuming the Yukawa parameter $y_N$ to be $O(1)$, it is straightforward to convert the constraints on lifetime previously obtained into ones in $\theta_\chi$, or equivalently in $\delta M$ as a function of $M_\chi$. They are shown in Fig.~\ref{fig:dM} (using the same color conventions as in Fig.~\ref{fig:tauX}), restricting the mass range such that $M_\chi<M_N$, adopting $M_N\sim2\times 10^{13}~$GeV as a ``vanilla seesaw'' value. It turns out that the mass mixing parameter $\delta M$ must satisfy approximately $\delta M\lesssim 2\times 10^{-17}/[M_X/(10^9~\mathrm{GeV})]^{0.5}$~GeV for $M_X\gtrsim 10^9$~GeV.

Such low values for $\delta M$ call for a UV completion of the model. For illustration towards this direction, we exemplify a setup that can generate, after the spontaneous breaking of a global $U(1)$ symmetry, the low-energy operator $(\delta M\overline{{\chi^\mathrm{c}}_{\mathrm{L}}}N_\mathrm{R}+\mathrm{h.c.})$ with tiny $\delta M$ values and simultaneously $\delta M\gg \delta m_\mathrm{D}$. The UV particle content is based on a set of four Weyl fermion pairs $\{\psi_i,\tilde{\psi}_i\}_{i=1\dots4}$ and two scalar fields $\phi_\chi$ and $\phi_N$, with the charge assignment under the $U(1)$ symmetry given in Table~\ref{tab:Q}. Using for a moment a two-component notation, the most general renormalizable Lagrangian that respects the symmetries reads as
\begin{multline}
    \mathcal{L}_\mathrm{UV}\supset -\bigg(\lambda_\chi\phi_\chi\overline{{\chi^\mathrm{c}}_{\mathrm{L}}}\chi_{\mathrm{R}}+\lambda_N\phi_N\overline{{N^\mathrm{c}}_{\mathrm{L}}}N_{\mathrm{R}}+\sum_iM\tilde{\psi}_i\psi_i\\
    +\lambda_{1N}\phi_N\overline{{N^\mathrm{c}}_{\mathrm{L}}}\psi_1+\lambda_{4\chi}\phi_N\tilde{\psi}_4\chi_\mathrm{R}+y_N\tilde{H}\overline{L}N_\mathrm{R}+V(\phi_N)+V(\phi_\chi)\\
    +\lambda_{12\chi}\phi_\chi\tilde{\psi}_1\psi_2+\lambda_{23N}\phi_N\tilde{\psi}_2\psi_3+\lambda_{34N}\phi_N\tilde{\psi}_3\psi_4\bigg),
\end{multline}
with, for simplicity, $M$ a mass common to all Weyl fermions and close to the Planck scale $M_\mathrm{P}$, and $\lambda_i$ coupling constants. The introduction of two scalar fields allows for generating two different high-scale masses for $N_\mathrm{R}$ and $\chi_\mathrm{R}$ after the breaking of symmetry. Integrating out the superheavy fermions leads to the following effective Lagrangian, back in four-component notation,
\begin{multline}
    \mathcal{L}_\mathrm{eff}\supset -\bigg(\lambda_\chi\phi_\chi\overline{{\chi^\mathrm{c}}_{\mathrm{L}}}\chi_{\mathrm{R}}+\lambda_N\phi_N\overline{{N^\mathrm{c}}_{\mathrm{L}}}N_{\mathrm{R}}+y_N\tilde{H}\overline{L}N_\mathrm{R}\\
    +\frac{\lambda_{1N}\lambda_{12\chi}\lambda_{23N}\lambda_{34N}\lambda_{4\chi}}{M^4}\phi_N^4\phi_\chi\overline{{\chi^\mathrm{c}}_{\mathrm{L}}}N_\mathrm{R}+V(\phi_N)+V(\phi_\chi)\bigg),
\end{multline}
which, after spontaneous breaking of the high-scale symmetry, yields the low-energy Lagrangian $\mathcal{L}_\mathrm{M}+\lambda_\mathrm{L}\tilde{H}\overline{L}N_\mathrm{R}$ with the substitutions
\begin{eqnarray}
    M_\chi&=&2\lambda_\chi\langle\phi_\chi\rangle ,\\
    M_N&=& 2\lambda_N\langle\phi_N\rangle,\\
    \delta M&=&\frac{\lambda_{1N}\lambda_{12\chi}\lambda_{23N}\lambda_{34N}\lambda_{4\chi}}{M^4}\langle\phi_N\rangle^4\langle\phi_\chi\rangle.
\end{eqnarray}
With $\langle\phi_N\rangle\sim[10^{13}-10^{14}]~$GeV and $\lambda_i$ values ranging between $O(0.01)$ and  $O(0.1)$, we see that the required orders of magnitude are obtained for $\delta M$ with $\langle\phi_\chi\rangle$ ranging between $\gtrsim10^{9}~$GeV and $\lesssim10^{13}~$GeV. Note that, while forbidden by symmetries at high scale, an effective interaction of the type $y_\chi\tilde{H}\overline{L}\chi_\mathrm{R}$ is generated through the chain of interactions of nearest-neighbor superheavy fermions,
\begin{equation}
    \mathcal{L}_\mathrm{eff}\supset -\frac{y_N\lambda_{1N}\lambda_{12\chi}\lambda_{23N}\lambda_{34N}\lambda_{4\chi}}{M_NM^4}\phi_N^4\phi_\chi\tilde{H}\overline{L}\chi_\mathrm{R},
\end{equation}
which leads however to a Dirac mass term $\delta m_\mathrm{D}\ll \delta M$ after symmetry breaking. This justifies the hierarchy assumed in Section~\ref{sec:seesawmodel}.

\subsection{Cosmological viability}
\label{sec:cosmology}

To complete the model, it remains to explore the viable range of cosmological parameters that guarantee the production of the DM candidate in the early Universe in adequate quantities to match the relic density observed today. We restrict ourselves to the minimal and unavoidable gravitational production, therefore without invoking fine-tuning couplings between the candidate and the inflaton. 

Cosmological gravitational particle production from the vacuum results from a rapid time evolution of the background gravitational field, see e.g.~\cite{Lyth:1996yj,Chung:1998zb,Kuzmin:1999zk,Chung:2001cb} for seminal studies in the context of DM as well as e.g.~\cite{Ema:2018ucl,Chung:2018ayg,Ema:2019yrd,Ahmed:2020fhc,Gross:2020zam,Kolb:2020fwh,Kaneta:2022gug} for recent studies. The inflationary era provides a gravitational background that varies so rapidly that it constitutes a portal to any sector of spin $1/2$ particles. This portal is efficient on the condition that particles are superheavy, as chiral symmetries guarantee scale-invariant couplings to the metric in the massless limit, in which case effects of the time-dependent metric can be eliminated. Note that, unlike many DM production scenarios, this mechanism offers the possibility of producing particles whose mass is well above the highest temperature reached during the reheating era; what matters most is the Hubble rate $H_\mathrm{end}$ at the end of inflation. 

\begin{figure}[t]
\centering
\includegraphics[width=0.5\textwidth]{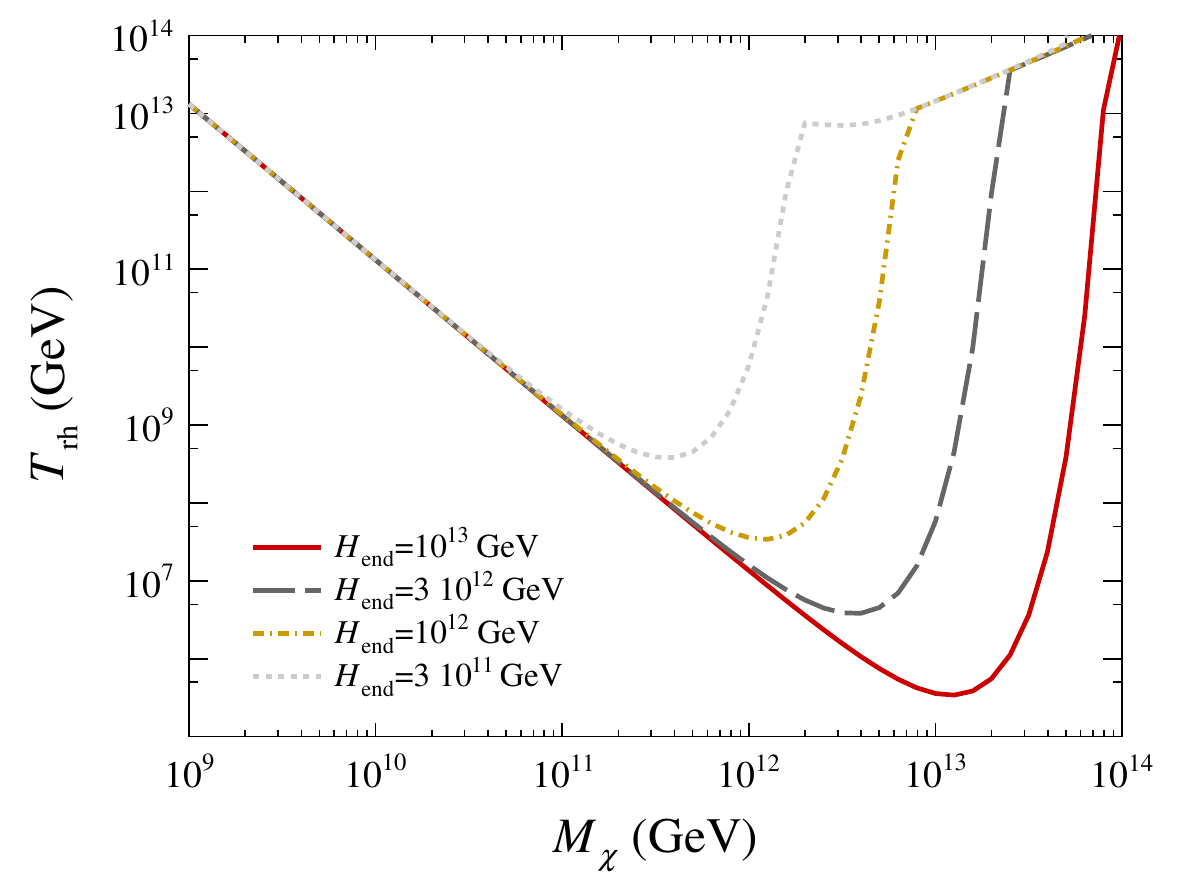}
\caption{Viable range of reheating temperature as a function of $M_\chi$ for cosmological gravitational particle production or gravitational freeze-in to be responsible for DM creation.}
\label{fig:Trh}
\end{figure}

For fermions produced gravitationally by the end of inflation, the dimensionless relic abundance of DM observed today, $\Omega_\chi h^2$, expressed from the energy density of DM, $\rho_\chi$, relative to the critical energy, $\rho_\mathrm{c}=3.61\times 10^{-47}~h^2~\mathrm{GeV}^{4}$, can be summarized in the following compact manner~\cite{Kolb:2023ydq}, 
\begin{equation}
    \label{eqn:OmegaX}
    \frac{\Omega_\chi h^2}{0.12}=\frac{10^5}{16\pi^2}\left(\frac{M_\chi}{H_\mathrm{end}}\right)^2\left(\frac{H_\mathrm{end}}{10^{12}~\mathrm{GeV}}\right)^2\left(\frac{T_\mathrm{rh}}{10^{9}~\mathrm{GeV}}\right)\mathrm{e}^{-\left(\frac{M_\chi}{H_\mathrm{end}}\right)^{3/2}},
\end{equation}
with $T_\mathrm{rh}$ the reheating temperature,  $h=H_0/(100~\mathrm{km/s/Mpc})$ the expansion rate, and $H_0$ the Hubble rate today. Note that this parameterization captures approximately the particle creation not only from the inflation era but also from the reheating one, when the rate of inflaton annihilations via an $s$-channel is enhanced~\cite{Ema:2019yrd,Mambrini:2021zpp,Barman:2021ugy,Clery:2021bwz,Clery:2022wib}. This is because the scale factor develops an oscillatory component that follows the inflaton field oscillations~\cite{Kaneta:2022gug}. Limiting ourselves to single-field slow-roll inflation, the parameter $H_\mathrm{end}$ can be related to the limits on the ratio $r$ of the tensor-to-scalar metric perturbations in the cosmic microwave background through~\cite{Lyth:1998xn,Planck:2018jri}
\begin{equation}
\label{eqn:r}
    H_\mathrm{end}=1.06\times 10^{-4}M_\mathrm{P}\sqrt{r}.
\end{equation}
Combining data from \textit{Planck}, BICEP2, \textit{Keck Array} and BICEP3 leads to $r<0.036$ at 95\% confidence level~\cite{BICEP:2021xfz}. This translates into 
\begin{equation}
    H_\mathrm{end}<4.8\times10^{13}~\mathrm{GeV}.
\end{equation}
We show in Fig.~\ref{fig:Trh} the viable range of reheating temperature $T_\mathrm{rh}$ as a function of $M_\chi$ for $H_\mathrm{end}=10^{13}~$GeV (red curve), $H_\mathrm{end}=3\times 10^{12}~$GeV (gray long-dotted curve), $H_\mathrm{end}=10^{12}~$GeV (orange dashed-dotted curve), and $H_\mathrm{end}=3\times 10^{12}~$GeV (light gray dotted curve). As anticipated, the gravitational production of $\tilde{\chi}$ is most efficient for $M_\chi\simeq H_\mathrm{end}$, where the reheating temperature is the lowest. The behavior of $T_\mathrm{rh}$ at low and high values in the mass range considered translates that the gravitational spin-$1/2$ particle production is inefficient at low $M_\chi$ (growing as $M_\chi^2$), and exponentially suppressed for $M_\chi>H_\mathrm{end}$. The flattening of $T_\mathrm{rh}$ at high $M_\chi$ values is due to another production mechanism, described below.

Besides cosmological gravitational particle production, the DM candidate can also be produced by ``gravitational freeze-in''~\cite{Garny:2015sjg,Garny:2017kha,Tang:2017hvq,Chianese:2020khl,Chianese:2020yjo,Redi:2020ffc,Bernal:2018qlk}, i.e.~by freeze-in mechanism~\cite{McDonald:2001vt,Choi:2005vq,Shaposhnikov:2006xi,Kusenko:2006rh,Petraki:2007gq,Hall:2009bx,Bernal:2017kxu} induced by graviton exchanges alone. In the framework of particle creation from the thermal bath governed by a Boltzmann equation, the DM yield $Y_\chi(a)$ expressed as the comoving density number $a^3n_\chi(a)$ relative to the entropy at reheating $s_\mathrm{rh}\equiv(a_\mathrm{rh})$ reads as~\cite{Garny:2015sjg}
\begin{equation}
    \label{eqn:Y}
    Y_\chi(a_\mathrm{rh})=\frac{1}{s_\mathrm{rh}}\int_{a_\mathrm{end}}^{a_\mathrm{rh}}\dif a \frac{a^2R(a)}{H(a)},
\end{equation}
with $a$ the scale factor, $H(a)$ the Hubble rate, and $R(a)$ the rate of annihilations of the bath particles into DM. For an $s$-channel, this latter is known to be, in the limit of massless particles in the incoming state~\cite{Gondolo:1990dk},
\begin{equation}
    \label{eqn:R}
    R(a)=\sum_i\frac{g_i^2T(a)}{2(2\pi)^4}\int\dif s~ 
    s^{3/2}  K_1\left(\frac{\sqrt{s}}{T(a)}\right)\sigma_i(s)
\end{equation}
with the sum running over all Standard Model degrees of freedom, $g_i$ the number of spin states, $T(a)$ the temperature,  $K_1$ the modified Bessel function of the second kind with order 1, and $\sigma_i(s)$ the cross section describing the $ii\rightarrow\tilde{\chi}\tilde{\chi}$ process via graviton exchanges, which, for Majorana $\tilde{\chi}$ particles, reads as
\begin{equation}
    \label{eqn:sigma}
    \sigma_i(s)=\frac{1}{64\pi sg_i^2}\frac{|\mathbf{p}_\chi|}{|\mathbf{p}_i|}\int\dif \cos\theta~\left|\mathcal{M}\right|^2.
\end{equation}
Here, $\mathbf{p}_\chi$ and $\mathbf{p}_i$ are the 3-momenta of the particles and the squared matrix element $\mathcal{M}$ is summed over all polarization states -- taken from~\cite{Tang:2017hvq}. Note that the $s$-channel $\tilde{\nu}\tilde{\nu}\rightarrow\tilde{\nu}\tilde{\chi}$ via $h$ exchange, as allowed by the model, yields a totally negligible production rate given the constraints on $\theta_\chi$ derived in previous sections. The dimensionless abundance of DM observed today is finally estimated as
\begin{equation}
    \label{eqn:OmegaXbis}
    \Omega_\chi h^2=\frac{M_\chi s_0Y_\chi(a_\mathrm{rh})}{a_\mathrm{rh}^3},
\end{equation}
with $s_0\simeq 2.23\times10^{-38}~$GeV$^3$ the entropy today. Using the reheating dynamics that governs the evolution of $H(a)$ given a value for $T_\mathrm{rh}$ and for $H_\mathrm{end}$, it is straightforward to infer $Y_\chi(a_\mathrm{rh})$ from Eqn.~\ref{eqn:Y} and subsequently the viable curves for $T_\mathrm{rh}$ as a function of $M_\chi$ via Eqn.~\ref{eqn:r} under the constrain $\Omega_\chi h^2=0.12$. They are shown in Fig.~\ref{fig:Trh} at high $M_\chi$ values, when this production mechanism dominates over the previous one. As expected, the reheating temperatures must be high, above $M_\chi$, for this mechanism to be efficient. 

Overall, there are wide ranges of viable values of cosmological parameters for which the right amount of DM can be obtained from gravitational particle production alone, be it from the rapid variations of the metric or from annihilation of the bath of particles during reheating.

\section{Conclusion}
\label{sec:discussion}

The prompt emission of ultrahigh energy gamma rays and neutrinos subsequent to the decay of superheavy particles in the Galaxy is known to provide a clean signature for a serendipitous discovery of superheavy DM. We have shown that, in addition, the secondary emission of ultrahigh energy gamma rays from the synchrotron radiation of electron byproducts is also constraining, being dominant over the prompt emission for DM candidates with masses $\gtrsim 10^{13}~$GeV. The non-observation, currently, of such ultrahigh energy gamma rays provides stringent limits on the lifetime of superheavy DM. 

These limits are powerful to constrain specific models of superheavy DM. As a concrete example, we considered a setup with chiral counterparts to left-handed neutrinos. In addition to generating the mass of light neutrinos through the seesaw mechanism, and despite the absence of direct coupling between one right-handed degree of freedom and the massless left-handed neutrino, we have shown that such an extended type-I seesaw framework can generate a viable DM candidate through a tiny mass-mixing parameter, which is constrained by the bounds on the lifetime and which can be generated by the breaking of a global symmetry at UV scale. Overall, neutrino physics is a data driven subject. The still unknown sector of right-handed neutrinos may be key to describe simultaneously the mass of the light neutrinos, superheavy DM and the baryon asymmetry in the Universe~\cite{Kuzmin:1985mm,Fukugita:1986hr}.



\bibliographystyle{epjc} 
\bibliography{biblio}

\end{document}